\documentclass[12pt]{iopart}

\newcommand{\om}{\omega}
\newcommand{\al}{\alpha}

\newcommand{\lm}{\lambda}

\newcommand{\half}{\frac{1}{2}}
\newcommand{\ie}{\textit{i.e.}}
\newcommand{\binomial}[2]{{#1 \choose #2}}
\begin{document}

\title[SU(2) Coherent states and Lissajous orbits]{Commensurate
 anisotropic oscillator, SU(2) coherent states and the classical limit}

\author{
M. Sanjay Kumar$^{1}$ and B. Dutta-Roy$^{2}$
}
\address{$^1$ S.N. Bose National Centre for Basic Sciences,
JD Block, Sector III, Salt Lake, Kolkata - 700098, India}
\address{$^{2}$ H-19 Baishnabghata Patuli, Kolkata - 700 094, India}
\ead{sanjay@bose.res.in}

\begin{abstract}
We demonstrate a formally exact quantum-classical correspondence
between the stationary coherent states
associated with the commensurate anisotropic two-dimensional harmonic
oscillator and the classical Lissajous orbits.
Our derivation draws upon earlier work of
Louck \etal [1973 \textit {J. Math. Phys.}
\textbf {14} 692] wherein they have provided a
non-bijective canonical transformation that maps, within a degenerate
eigenspace, the
commensurate anisotropic oscillator on to the isotropic oscillator.
This mapping leads, in a natural manner, to a Schwinger realization
of $SU(2)$ in terms of the canonically transformed creation and
annihilation operators.
Through the corresponding coherent states built over a degenerate
eigenspace, we directly effect the classical limit via the expectation
values of the underlying generators. Our work completely accounts for
the fact that the SU(2) coherent state in general corresponds to an
ensemble of Lissajous orbits.
\end{abstract}





\section{Introduction}
\label{Sec:Intro}

The anisotropic oscillator has long been of relevance in describing
the intrinsic states of a deformed nucleus in the Nilsson
Model \cite{nilsson}. The discovery of super-deformed high spin states
of some nuclei \cite{stephens} corresponding to spheroidal nuclear
shapes of approximately commensurate axial lengths had helped focus
attention on the commensurate anisotropic oscillator. 
Similarly in
quantum optics a two-mode radiation field may also be discussed in
terms of a two-dimensional oscillator \cite{agarwal-vortex}. Likewise
in condensed matter physics, the design of nanostructures permitting
ballistic motion of electrons \cite{reed,kelly} represents yet another
area for the
application of such studies.

Considerable attention has been paid in the literature on the
question of symmetries
and degeneracies in the commensurate anisotropic
oscillator \cite{jauch-hill,king,louck,rosensteel,bhaumik}.
In particular Louck \etal have addressed this question from
a group theoretical viewpoint by studying the non-bijective
canonical transformation that maps the commensurate anisotropic
oscillator, within a degenerate eigenspace, to the isotropic one.

While the question of achieving the classical limit of quantum
dynamics of simple systems via appropriately constructed coherent states
\cite{schrodinger,klauder-skagerstam,chaturvedi-agarwal,ingold} has been a
long-standing one, interesting experiments have been carried
out recently to demonstrate such a classical limit in quantum
systems \cite{stroud}. More recently
the classical limit of the commensurate
anisotropic oscillator has been investigated experimentally
\cite{chen-expt} in a laser resonator by exploiting the analogy
between the Schr\"odinger equation for the two-dimensional harmonic
oscillator and the paraxial wave equation for the spherical
resonators \cite{haus,siegman}. The question of analytically demonstrating the
classical limit in this system via appropriately constructed
coherent states and accounting for the experimentally
observed wave patterns has been an intriguing one and has been
addressed by various
authors \cite{chen-expt,chen-huang1,makowski,chen-nonlin,gorska}. The purpose
of this paper is to resolve this question using an approach that
exploits the symmetry properties of the commensurate two-dimensional
anisotropic oscillator well studied in the literature \cite{louck}.

Consider the two-dimensional harmonic oscillator described by the
Hamiltonian \cite{chen-nonlin},

\begin{equation}
H = \half \left(\hat{p}_{x}^{2}+\hat{p}_{y}^{2}+
\om_{1}^{2}\hat{x}^{2}+ \om_{2}^{2}\hat{y}^{2} \right),
\label{eq:1}
\end{equation}
where $\om_{1}=q\om$ and $\om_{2}=p\om$, $\om$ is the common factor of
the frequencies $\om_{1}$ and $\om_{2}$, and $p$ and $q$ are
integers. Normally one takes $p$ and $q$ to be coprime, without loss
of generality, as the common factor between $p$ and $q$ ($M$ say) can
be absorbed in the definition of the common frequency $\om$. However,
we take the Hamiltonian \eref{eq:1} here to describe the experimental
situation of Chen \etal \cite{chen-expt} where the common frequency $\om$
represents the tranverse mode spacing in the spherical resonator, and
$p$ and $q$ can be independently varied
by suitably tuning the cavity length and appropriately choosing the
longitudinal mode indices. Thus $p$ and $q$ could in practice have
a common factor $M\neq 1$. Further, as reported
by Chen \etal \cite{chen-expt}, the experimental situations
corresponding to the choice of parameters ($p,q$) and ($lp,lq$) where
$l$ is a positive integer, give rise to qualitatively different
results in regard to quantum-classical correspondence. In view of
this, in the rest of the paper, we take $p$ and $q$ to be having a
common factor $M$ in general.

The Hamiltonian \eref{eq:1} can be written in terms of the creation
and annihilation operators in the form,
\begin{equation} H =
\om^{\prime}\left[ \frac{1}{p}\left(a_{1}^{\dag}a_{1} + \half \right)+
\frac{1}{q}\left(a_{2}^{\dag}a_{2} + \half \right) \right],
\label{eq:2}
\end{equation}
where $\om^{\prime}=\om pq$ and
\begin{equation} a_{1} =
\frac{1}{\sqrt{2q\om}}\left(q\om\hat{x}+i\hat{p}_{x}\right), \qquad
a_{2} = \frac{1}{\sqrt{2p\om}}\left(p\om\hat{y}+i\hat{p}_{y}\right).
\label{eq:3}
\end{equation}

It is in fact straightforward to achieve the classical limit of the
quantum dynamics described by the Hamiltonian \eref{eq:2} via the
two-mode harmonic oscillator coherent states $\vert
\al_{1},\al_{2}\rangle$ that are defined by
\begin{equation} a_{1}
\vert \al_{1},\al_{2}\rangle = \al_{1} \vert \al_{1},\al_{2}\rangle,
\qquad a_{2} \vert \al_{1},\al_{2}\rangle = \al_{2} \vert
\al_{1},\al_{2}\rangle.
\label{eq:4}
\end{equation} Note that these
are the coherent states associated with the Heisenberg-Weyl group
\cite{perelomov}.  Let the system be initially (at $t=0$) in the
two-mode coherent state $\vert \al_{1},\al_{2}\rangle$.  The
expectation values of $a_{1}$, $a_{2}$ in this state evolve in time
under the Hamiltonian \eref{eq:2} as
\begin{equation}
\langle a_{1}(t)\rangle = \al_{1} e^{-iq\om t}, \qquad \langle a_{2}(t)\rangle
= \al_{2} e^{-ip\om t}. 
\label{eq:5}
\end{equation}
The classical
Hamiltonian corresponding to \eref{eq:1} can be rewritten in the form
\begin{equation}
H = \om^{\prime}\left( \frac{1}{p}\vert
z_{1}\vert^{2}+ \frac{1}{q}\vert z_{2}\vert^{2} \right),
\label{eq:6}
\end{equation}
where the complex variables ($z_{1},z_{2}$) are related
to the classical coordinates $x$, $y$ and momenta $p_{x}$, $p_{y}$ by
\begin{equation}
z_{1} = \frac{1}{\sqrt{2q\om}}\left(q\om
x+ip_{x}\right), \qquad z_{2} = \frac{1}{\sqrt{2p\om}}\left(p\om
y+ip_{y}\right),
\label{eq:7}
\end{equation}
with $\om^{\prime}=\om
pq$ as defined earlier.  Let us write the solutions of the classical
Hamiltonian \eref{eq:6} as
\begin{equation} z_{1}(t) = \sqrt{\frac{\om
q}{2}} \eta_{1} e^{-i(\om qt-\phi_{1})}, \qquad z_{2}(t) =
\sqrt{\frac{\om p}{2}} \eta_{2} e^{-i(\om pt-\phi_{2})},
\label{eq:8}
\end{equation}
so that the equations describing the classical
Lissajous orbits would be given by
\begin{equation}
x(t) = \eta_{1}\cos(q\om t-\phi_{1}), \qquad y(t) = \eta_{2}\cos(p\om
t-\phi_{2}). 
\label{eq:9}
\end{equation}

The position probability density, namely, $\vert\langle
x,y\vert \al_{1},\al_{2}\rangle \vert^{2}$ is Gaussian centred at
[$(\langle a_{1}\rangle$+$\langle a_{1}^{\dag} \rangle)/\sqrt{2q\om}$,
$(\langle a_{2} \rangle$+$\langle a_{2}^{\dag} \rangle)/\sqrt{2p\om}$],
and becomes localized
at this point in the classical limit, \ie, $\hbar \rightarrow 0$.
Thus as time evolves the peak of the position probability density,
rides on the classical trajectory \eref{eq:9}.  This suggests the
following prescription for implementing the classical limit: the
expectation values of the generators $a_{1}$, $a_{1}^{\dag}$, $a_{2}$,
$a_{2}^{\dag}$, of the Heisenberg-Weyl group, in the two-mode coherent
state, tend to the corresponding classical values.  Thus the classical
limit in this case is obtained simply by making the correspondence
\begin{equation}
(\langle a_{1} \rangle, \langle a_{1}^{\dag} \rangle,
\langle a_{2} \rangle, \langle a_{2}^{\dag} \rangle) \longrightarrow
(z_{1},z_{1}^{*}, z_{2},z_{2}^{*}). 
\label{eq:10}
\end{equation}
The above correspondence is also evident from the the formal similarity
between the solutions \eref{eq:5} for the expectation values and the
solutions \eref{eq:8} for the corresponding classical dynamical
variables.  This correspondence yields a relation between the
parameters in the equations for the Lissajous orbits \eref{eq:9} and
the coherent state $\vert \al_{1},\al_{2}\rangle$ as
\begin{equation}
\al_{1} = \sqrt{\frac{\om q}{2}} \eta_{1} e^{i\phi_{1}}, \qquad
\al_{2} = \sqrt{\frac{\om p}{2}} \eta_{2} e^{i\phi_{2}}.
\label{eq:11}
\end{equation}
Thus there is a unique classical
trajectory corresponding to a given two-mode coherent state.

Note that the demonstration of the classical limit of the
two-dimensional oscillator that we have presented above, via the
two-mode coherent state $\vert \al_{1},\al_{2}\rangle$, would be valid
even if one considers the two frequencies $\om_{1}$, $\om_{2}$ to be
incommensurate.  This in fact is an unsatisfactory feature since the
coherent state $\vert \al_{1},\al_{2}\rangle$ does not embody the full
symmetry of the commensurate anisotropic oscillator Hamiltonian. To
illustrate this point, let us look at the special case of the
isotropic oscillator ($p=q=1$).  The $SU(2)$ symmetry in this case is
manifest as the classical Hamiltonian \eref{eq:6} preserves the form
$|z_{1}|^{2}+|z_{2}|^{2}$.  The quantum Hamiltonian \eref{eq:2} on the
other hand can be rewritten as
\begin{equation}
H = \om^{\prime}(2J_{0}+1),
\label{eq:12}
\end{equation}
where $J_{0}$ is
the Casimir operator corresponding to the $SU(2)$ Lie algebra
generated, in the Schwinger realization, by
\begin{equation}
\fl J_{+}
= a_{1}^{\dag} a_{2}, \qquad J_{-} = a_{1} a_{2}^{\dag}, \qquad J_{z}
= (a_{1}^{\dag}a_{1} - a_{2}^{\dag}a_{2})/2, \qquad J_{0} =
(a_{1}^{\dag}a_{1} + a_{2}^{\dag}a_{2})/2.  \label{eq:13}
\end{equation} Here the operators $J_{\pm}$, $J_{z}$ obey the standard
commutation relations \begin{equation} [ J_{z}, J_{\pm} ] = \pm
J_{\pm}, \qquad [ J_{+}, J_{-} ] = 2 J_{z}. 
\label{eq:14}
\end{equation}

In view of \eref{eq:13}, the set of simultaneous eigenstates
of $J^{2}=J_{0}(J_{0}+1)$ and $J_{z}$, namely $\vert j,m \rangle$,
$j=0,1,\ldots, \infty$, $|m| \leq j$, where $j=\half (n_{1}+n_{2})$ and
$m=\half (n_{1}-n_{2})$ is isomorphic to the set of number states
$\vert n_{1},n_{2}\rangle$. The isotropic oscillator Hamiltonian
divides this set of states into degenerate eigenspaces each characterized
$j$ (the eigenvalue of the Casimir operator $J_{0}$) independent of $m$.
The two-mode
harmonic oscillator coherent state $\vert \al_{1},\al_{2}\rangle$ can
then be expressed as \cite{takahashi-shibata}
\begin{equation}
\vert \al_{1},\al_{2}\rangle = \sum_{j=0}^{\infty}
e^{-\half(|\al_{1}|^{2}+|\al_{2}|^{2})}
(|\al_{1}|^{2}+|\al_{2}|^{2})^{j}
\left(\frac{\al_{2}}{|\al_{2}|}\right)^{j} \vert j,\tau \rangle,
\label{eq:15}
\end{equation}
where $\vert j,\tau \rangle$ is the
$SU(2)$ coherent state \cite{radcliffe,arecchi,perelomov} built over
states in the degenerate eigenspace \{$\vert j,m \rangle, \vert m
\vert \leq j$\} (equivalently the number states $\vert n_{1},n_{2}
\rangle$ with $n_{1}+n_{2}$ held fixed) namely,
\begin{equation}
\vert j,\tau \rangle = \frac{1}{(1+\vert \tau \vert ^{2})^{j}}
\sum_{m=-j}^{j} \binomial{2j}{j+m}^{\half} \tau^{j+m} \vert j,m
\rangle ,
\label{eq:16}
\end{equation}
with $\tau=\al_{1}/\al_{2}$.
Note that the two-mode coherent state $\vert \al_{1},\al_{2}\rangle$
involves a sum over all degenerate eigenspaces labelled by $j$, and
hence it does not implement the $SU(2)$ symmetry of the isotropic
oscillator Hamiltonian. The coherent state $\vert j,\tau \rangle$ on
the other hand, being a projection of the two-mode coherent state on
to a particular degenerate eigenspace characterized by the energy
$E=\om^{\prime}(2j+1)$, does respect this symmetry. In this sense the
appropriate coherent state which must be used to examine the classical
limit in the isotropic oscillator case is the $SU(2)$ coherent state
\eref{eq:16}.

Indeed, Bi\`evre  \cite{bievre} and Pollet \etal  \cite{pollet} have
used the $SU(2)$ coherent states to demonstrate the classical limit in
the case of the isotropic harmonic oscillator. In particular they have
rigorously demonstrated that the coordinate space probability density
$|\langle x,y \vert j,\tau\rangle |^{2}$ in the limit $2j=N
\rightarrow \infty$ becomes localized over the classical Lissajous
(elliptic) orbits.

More recently, the question of how to analytically derive a connection
between a suitably constructed coherent state for the commensurate
two-dimensional anisotropic oscillator and the classical Lissajous
orbits has acquired interest
\cite{chen-huang1,makowski,chen-nonlin,gorska}, especially with a view
to theoretically account for the experimental demonstration of such a
classical limit by Chen \etal \cite{chen-expt}.  Chen and coworkers
\cite{chen-expt,chen-huang1} have made an ansatz on the appropriate
coherent state, something that resembles an $SU(2)$ coherent state,
namely,
\begin{equation}
\vert N, p, q, \tau \rangle =
\frac{1}{(1+\vert \tau \vert ^{2})^{N/2}} \sum_{K=0}^{N}
\binomial{N}{K}^{\half} \tau^{K} \vert pK,q(N-K) \rangle ,
\label{eq:17}
\end{equation}
where $N$ is a non-negative integer. From
a numerical study of the coordinate space probability density
associated with the above state they have guessed the following
quantum-classical connection. For coprime $p$ and $q$ the classical
periodic orbit is given by
\begin{equation} x(t) = \eta_{1}\cos(q\om
t-\phi/p), \qquad y(t) = \eta_{2}\cos(p\om t),
\label{eq:18}
\end{equation}
with the amplitudes $\eta_{1}$ and $\eta_{2}$ given by
\begin{equation}
\eta_{1} = \sqrt{\frac{1}{\om q} \left(\frac{2pN\vert
\tau \vert^{2}}{1+\vert \tau \vert^{2}}+1\right)}, \qquad \eta_{2} =
\sqrt{\frac{1}{\om p} \left(\frac{2qN}{1+\vert \tau
\vert^{2}}+1\right)}, 
\label{eq:19}
\end{equation}
where $\phi$ is an arbitrary phase.  On the other hand
if $p$ and $q$ have a common factor $M$, the coordinate space
probability density is found to correspond to an ensemble of classical
periodic orbits, the total number of such periodic orbits being $M$,
and their trajectories are given by
\begin{eqnarray}
x_{k}(t) & = &
\eta_{1}\cos[q\om t-(\phi+2\pi k)/p], \qquad k=0,1,\ldots, M-1,
\nonumber \\
y(t) & = & \eta_{2}\cos(p\om t),
\label{eq:20}
\end{eqnarray} with $\eta_{1}$, $\eta_{2}$ as defined in \eref{eq:19}.

Chen \etal \cite{chen-nonlin} have, for the first time, attempted to
give an analytical derivation of the above guessed equations for the
classical periodic orbits in the commensurate anisotropic oscillator
case.  They effect the classical limit via the two-mode coherent state
$\vert \al_{1},\al_{2}\rangle$ as demonstrated above in equations
\eref{eq:10} and \eref{eq:11}, and then utilize the method of
triangular partial sums to essentially project a stationary `coherent'
state out of the two-mode coherent state $\vert
\al_{1},\al_{2}\rangle$.  They indicate a connection between the
parameters of this stationary coherent state and the classical
periodic orbits in the case when $p$ and $q$ are coprime, leaving the
question of what happens in the case when $p$ and $q$ have a common
factor $M\neq1$ unanswered. Unfortunately, their derivation does not
clearly bring out the fact that the coherent state that they have
projected out is indeed the $SU(2)$ coherent state. Not surprisingly
these authors have referred to the stationary state constructed by
them as \textit{a kind of} $SU(2)$ coherent state. G\'orska \etal
\cite{gorska} on the other hand offer an approximate correspondence
between the experimentally observed wave patterns and the classical
Lissajous orbits.

It is natural to expect, as in the isotropic oscillator case, that the
appropriate coherent states for the commensurate anisotropic
oscillator should be those associated with its underlying symmetry
group.  While the underlying $SU(2)$ group structure of the isotropic
oscillator is manifest, as outlined above, the fact that the group
$SU(2)$ also captures the symmetry of the two-dimensional commensurate
anisotropic oscillator has been shown by Louck \etal \cite{louck}.  In
particular they have concentrated on the degenerate eigenspaces of the
commensurate anisotropic oscillator and have constructed a
non-bijective canonical transformation that maps, within a degenerate
eigenspace, the commensurate anisotropic oscillator Hamiltonian to an
isotropic one, thus revealing the $SU(2)$ symmetry and also accounting
for the `accidental' degeneracy in the former case.  Furthermore, they
have also noted that this mapping leads, in a natural manner, to a
Schwinger realization of $SU(2)$ in terms of the canonically
transformed creation and annihilation operators, within a given
degenerate eigenspace.

In the present paper we use symmetry arguments to identify the
appropriate coherent states for the commensurate anisotropic
oscillator.  We utilize the above-mentioned canonical transformation
of Louck \etal, and the Schwinger realization of $SU(2)$ to construct
the stationary coherent states built over a degenerate subspace.  We
use these coherent states and demonstrate a correspondence with the
classical Lissajous orbits. In particular we derive a relation between
the parameters characterizing the $SU(2)$ coherent state and those
characterizing the single Lissajous orbit in the case when $p$ and $q$
are coprime, and an ensemble of $M$ Lissajous orbits when $p$ and $q$
have a common factor $M$.

\section{Canonical transformations and the symmetry group of the
commensurate anisotropic oscillator} \label{Sec:Canonical}

In this section we collect the main results from the work of Louck
\etal \cite{louck} that we shall make use of in the next section.  As
has been shown by Louck \etal \cite{louck}, the eigenstates of the
commensurate anisotropic oscillator Hamiltonian [with
$\om_{1}=q\om, \om_{2}=p\om$] can be divided into $qp$ number of different
subsets of states \cite{louck-footnote}
\begin{equation}
\left\{ \vert n_{1}p+\lm_{1},
n_{2}q+\lm_{2} \rangle, \quad n_{1},n_{2} = 0,1,\ldots, \infty\right\},
\label{eq:21}
\end{equation}
for each $\lm_{1} = 0,1,\ldots, p-1$, $\lm_{2} = 0,1,\ldots, q-1$.
The states in \eref{eq:21} are eigenstates of $H$ with eigenvalues
\begin{equation}
E = \om^{\prime}\left[(n_{1}+n_{2})+
\frac{1}{p}\left(\lm_{1}+\half\right)+
\frac{1}{q}\left(\lm_{2}+\half\right) \right],
\label{eq:22}
\end{equation}
so that those states belonging to the set \eref{eq:21} for a
fixed value of $n_{1}+n_{2}$ are degenerate.

In each of the degenerate eigenspaces \eref{eq:21} labelled by
($\lm_{1}$, $\lm_{2}$), there exists a canonical transformation
$(a_{1},a_{2})$ $\rightarrow$ $(\tilde{a}_{1},\tilde{a}_{2})$ given by
\begin{eqnarray}
\tilde{a}_{1} =&&
\sqrt{\frac{1}{p}(\hat{n}_{1}-\lm_{1})} \quad \hat{n}_{1}
(\hat{n}_{1}-1) (\hat{n}_{1}-p+1)^{-\half} (a_{1}^{\dag})^{p},
\nonumber \\
\tilde{a}_{2} =&& \sqrt{\frac{1}{q}(\hat{n}_{2}-\lm_{2})}
\quad \hat{n}_{2} (\hat{n}_{2}-1) (\hat{n}_{2}-q+1)^{-\half}
(a_{2}^{\dag})^{q}, 
\nonumber \\
\hat{n}_{1} =&& a_{1}^{\dag}a_{1},
\quad \hat{n}_{2} = a_{2}^{\dag}a_{2},
\label{eq:23}
\end{eqnarray}
such that the Hamiltonian in the transformed picture becomes that of
an isotropic oscillator with frequency $\om^{\prime}$, namely,
\begin{equation}
H = \om^{\prime}\left[
\left(\tilde{a}_{1}^{\dag}\tilde{a}_{1}+\half \right)+
\left(\tilde{a}_{2}^{\dag}\tilde{a}_{2}+\half \right) \right].
\label{eq:24}
\end{equation}

Note that the action of the canonically transformed creation and
annihilation operators on a particular state in the subset of states
\eref{eq:21} is given by, for example,
\begin{eqnarray}
\tilde{a}_{1}^{\dag} \vert n_{1}p+\lm_{1}, n_{2}q+\lm_{2}\rangle =&&
\sqrt{n_{1}+1} \vert (n_{1}+1)p+\lm_{1}, n_{2}q+\lm_{2}\rangle,
\nonumber \\
\tilde{a}_{2} \vert n_{1}p+\lm_{1}, n_{2}q+\lm_{2}\rangle
=&& \sqrt{n_{2}} \vert n_{1}p+\lm_{1}, (n_{2}-1)q+\lm_{2}\rangle,
\label{eq:25}
\end{eqnarray}
and so on.

As observed by Louck \etal \cite{louck}, one has the Schwinger
realization of $SU(2)$ in terms of the canonically transformed
operators $\tilde{a}_{1}$, $\tilde{a}_{1}^{\dag}$, $\tilde{a}_{2}$,
$\tilde{a}_{2}^{\dag}$, namely,
\begin{equation}
\fl J_{+} = \tilde{a}_{1}^{\dag} \tilde{a}_{2}, \qquad J_{-} = \tilde{a}_{1}
\tilde{a}_{2}^{\dag}, \qquad J_{z} =
(\tilde{a}_{1}^{\dag}\tilde{a}_{1} -
\tilde{a}_{2}^{\dag}\tilde{a}_{2})/2, \qquad J_{0} =
(\tilde{a}_{1}^{\dag}\tilde{a}_{1} +
\tilde{a}_{2}^{\dag}\tilde{a}_{2})/2,
\label{eq:26}
\end{equation}
where the operators $J_{\pm}$, $J_{z}$ obey the commutation relations
\eref{eq:14}.

In view of \eref{eq:25}, one can identify, for fixed
($\lm_{1},\lm_{2}$), the simultaneous eigenstates of
$J^{2}=J_{0}(J_{0}+1)$ and $J_{z}$, namely $\vert j,m \rangle$,
where $j=\half (n_{1}+n_{2})$ and $m=\half (n_{1}-n_{2})$,
with $\vert n_{1}p+\lm_{1}, n_{2}q+\lm_{2}\rangle$.  In terms of the
generators of $SU(2)$ defined in \eref{eq:26} the Hamiltonian
\eref{eq:24} is given by
\begin{equation}
H = \om^{\prime}(2J_{0}+1),
\label{eq:27}
\end{equation} so that for fixed ($\lm_{1},\lm_{2}$),
the energy eigenvalue in the state $\vert j,m \rangle$ is given by
$E=\om^{\prime}(2j+1)$ independent of $m$. This again reveals the `accidental'
degeneracy of the commensurate anisotropic oscillator due to the
underlying $SU(2)$ symmetry group.

We would like to recall here that Louck \etal \cite{louck} have also
provided a canonical transformation $(z_{1},z_{2})$ $\rightarrow$
$(\tilde{z}_{1},\tilde{z}_{2})$, in terms of complex variables defined
in \eref{eq:7}, given by
\begin{equation}
\tilde{z}_{1} =
\frac{1}{\sqrt{p}} \left(\frac{z_{1}}{\vert z_{1}\vert}\right)^{p}
\vert z_{1} \vert, \qquad \tilde{z}_{2} = \frac{1}{\sqrt{q}}
\left(\frac{z_{2}}{\vert z_{2}\vert}\right)^{q} \vert z_{2} \vert,
\label{eq:28}
\end{equation} such that the classical Hamiltonian
\eref{eq:6} in the transformed picture becomes that of the classical
isotropic harmonic oscillator, namely,
\begin{equation}
H = \om^{\prime}\left(
\vert \tilde{z}_{1}\vert^{2}+ \vert \tilde{z}_{2}\vert^{2} \right).
\label{eq:29}
\end{equation}
The $SU(2)$ symmetry of the classical
Hamiltonian \eref{eq:29} is evident again due to the fact that the
form $\vert \tilde{z}_{1}\vert^{2}+\vert \tilde{z}_{2}\vert^{2}$ is
preserved.

\section{$SU(2)$ coherent states, stereographic projection and
Lissajous orbits} \label{Sec:Lissajous}

Let us construct a $SU(2)$ coherent state out of states in the
degenerate eigenspace \{$\vert j,m \rangle, \vert m \vert \leq j$\}
for fixed ($\lm_{1},\lm_{2}$), namely,
\begin{eqnarray}
\vert j,\tau
\rangle & = & \frac{1}{(1+\vert \tau \vert ^{2})^{j}} \sum_{m=-j}^{j}
\binomial{2j}{j+m}^{\half} \tau^{j+m} \vert j,m \rangle, \nonumber \\
\tau & = & \tan \frac{\theta}{2}e^{i\phi},\quad 0 \leq \theta \leq
\pi, \quad 0 \leq \phi \leq 2\pi. 
\label{eq:30}
\end{eqnarray}

We would like to remark that, in view of the isomorphism between the
states $\vert n_{1}p+\lm_{1}, n_{2}q+\lm_{2}\rangle$ and the angular
momentum eigenstates $\vert j,m \rangle$, for fixed
($\lm_{1},\lm_{2}$), where $j=\half (n_{1}+n_{2})$ and $m=\half
(n_{1}-n_{2})$, one can see that the $SU(2)$ coherent state defined
above is equivalent to the `coherent' state \eref{eq:17} considered
earlier by Chen and coworkers
\cite{chen-expt,chen-huang1,chen-nonlin} if one makes the
identification $N=2j$ and specializes to ($\lm_{1},\lm_{2}$)=($0,0$).
Thus we have provided a symmetry-based justfication for the
particular form of the coherent state, that Chen and coworkers had
only conjectured based on heuristic considerations.  As will become evident
from the following analysis, the classical limit is independent of
the choice of $\lm_{1},\lm_{2}$, \ie, it does not matter which
degenerate eigenspace one works in.

Let the system be initially (at $t=0$) in the $SU(2)$ coherent state
as defined in \eref{eq:30}.  As time evolves the system remains in
the initial coherent state except for an irrelevant phase factor
$e^{-i\om^{\prime}t(2j+1)}$, so that the expectation values of
$J_{x}=(J_{+}+J_{-})/2$, $J_{y}=-i(J_{+}-J_{-})/2$, and $J_{z}$
remain stationary and are given by
\begin{equation}
\langle J_{x}\rangle = j\sin\theta \cos\phi, \qquad \langle J_{y}\rangle =
-j\sin\theta \sin\phi, \qquad \langle J_{z}\rangle = -j\cos\theta.
\label{eq:31}
\end{equation}
Clearly the point ($\langle
J_{x}\rangle, \langle J_{y}\rangle, \langle J_{z}\rangle$) lies on a
sphere of radius $j$.

Let us next consider the solutions of the classical Hamiltonian
\eref{eq:6} given in \eref{eq:8}.  In terms of the canonically
transformed complex variables ($\tilde{z}_{1},\tilde{z}_{2}$) defined
in \eref{eq:28} these solutions become
\begin{equation}
\tilde{z}_{1}(t) = \frac{1}{\sqrt{p}} \sqrt{\frac{\om q}{2}} \eta_{1}
e^{-ip(\om qt-\phi_{1})}, \qquad \tilde{z}_{2}(t) =
\frac{1}{\sqrt{q}} \sqrt{\frac{\om p}{2}} \eta_{2} e^{-iq(\om
pt-\phi_{2})}. 
\label{eq:32}
\end{equation}
Note that while the
solutions generated by the classical Hamiltonian \eref{eq:6}, namely
$z_{1}(t)$ and $z_{2}(t)$, oscillate at frequencies $\om p$ and $\om
q$ respectively, the solutions generated by the canonically
transformed Hamiltonian \eref{eq:20} oscillate at the common
frequency $\om pq$.  We would like to remark that the canonical
transformation \eref{eq:28} given by Louck \etal, although it is a
transformation of phase space variables, when regarded as a
trasformation among the coordinates alone, amounts to an
\textit{untwisting} of the Lissajous figures into a generic ellipse.

Since $\vert \tilde{z}_{1}\vert^{2} + \vert \tilde{z}_{2}\vert^{2}$
is a constant in view of energy conservation \eref{eq:29}, there
exists a mapping (stereographic projection) from a point
($j_{x},j_{y},j_{z}$) on a sphere of radius $j$, via the north pole,
to the complex $Z$-plane where we have defined $Z$ to be
\begin{equation}
Z = 2j \frac{\tilde{z}_{2}}{\tilde{z}_{1}}.
\label{eq:33}
\end{equation}
The stereographic projection from
($j_{x},j_{y},j_{z}$) to $Z$ is given by
\begin{equation}
Z = \frac{2j}{j-j_{z}} (j_{x}+ij_{y}). 
\label{eq:34}
\end{equation}

Recall that we have earlier effected the transition to the classical
limit \eref{eq:10} of the two-dimensional oscillator by identifying
the expectation values of the generators of the Heisenberg-Weyl
group, namely $a_{1}$, $a_{1}^{\dag}$, $a_{2}$, $a_{2}^{\dag}$, in
the two-mode coherent states, with the corresponding classical phase
space values. Motivated by this we prescribe that the transition to
the classical limit of the commensurate anisotropic two-dimensional
oscillator, in terms of the $SU(2)$ coherent state, can be effected
in analogy with \eref{eq:10}, by making the correspondence between
the expectation values of the generators of $SU(2)$ in the $SU(2)$
coherent states, namely,
\begin{equation}
(\langle J_{x}\rangle,
\langle J_{y}\rangle, \langle J_{z}\rangle) \longrightarrow (j_{x},
j_{y}, j_{z}),
\label{eq:35}
\end{equation}
where ($j_{x}, j_{y},
j_{z}$) is the point on the sphere of radius $j$ corresponding to the
pair of complex numbers ($\tilde{z}_{1},\tilde{z}_{2}$) that form the
solution set \eref{eq:32} of the classical isotropic oscillator
Hamiltonian in the transformed picture \eref{eq:29}. In fact such a
quantum-classical correspondence is implicit in the analysis of
Bi\`evre \cite{bievre} and Pollet \etal \cite{pollet} in the case of
the isotropic oscillator.

In view of the above proposed correspondence \eref{eq:35} we therefore
have
\begin{equation}
j_{x} = j\sin\theta \cos\phi, \qquad
j_{y} = -j\sin\theta \sin\phi, \qquad
j_{z} = -j\cos\theta,
\label{eq:36}
\end{equation}
and hence in view of \eref{eq:34} the complex variable $Z$
in the projective plane is related to the parameters in the $SU(2)$
coherent state \eref{eq:30} by
\begin{equation}
Z = 2j \cot\frac{\theta}{2}e^{-i\phi} = \frac{2j}{\tau}.
\label{eq:37}
\end{equation}
Upon combining this result with \eref{eq:32} and \eref{eq:33} we have
the relations
\begin{equation}
\frac{q\eta_{1}}{p\eta_{2}} = \vert \tau \vert
\label{eq:38}
\end{equation}
and
\begin{equation}
e^{i(p\phi_{1}-q\phi_{2})} = e^{i\phi}.
\label{eq:39}
\end{equation}

The relation \eref{eq:38}, in conjunction with the identification of
the classical expression of energy \eref{eq:6} with the eigenvalue of
the quantum Hamiltonian \eref{eq:27} in the $SU(2)$ coherent state,
namely,
\begin{equation}
\om^{2}\left(
\frac{q^{2}}{2}\eta_{1}^{2} +
\frac{p^{2}}{2}\eta_{2}^{2}
\right)
= \om^{\prime}(2j+1) = \om pq(N+1),
\label{eq:40}
\end{equation}
leads to the solutions for $\eta_{1}$ and $\eta_{2}$,
\begin{equation}
\eta_{1} = \sqrt{\frac{2p(N+1)}{q\om}}
\frac{\vert \tau \vert}{\sqrt{1+\vert \tau \vert^{2}}}, \qquad
\eta_{2} = \sqrt{\frac{2q(N+1)}{p\om}}
\frac{1}{\sqrt{1+\vert \tau \vert^{2}}}.
\label{eq:41}
\end{equation}

The solution of \eref{eq:39} needs detailed consideration.
The general solution of the relation \eref{eq:39} may be written as
\begin{equation}
p\phi_{1}-q\phi_{2} = \phi + 2\pi k,
\label{eq:42}
\end{equation}
where $k$ is an arbitrary integer. We shall now try to fix the allowed
range of values of $k$. As we shall see this will depend on whether
$p$ and $q$ are coprime or not.
Note that keeping $\phi_{1}$ fixed for example while varying $\phi_{2}$
in equation \eref{eq:9}
would only change the initial point on the Lissajous orbit and hence
would leave
the shape of the orbit itself invariant. This reparametrization
invariance of the Lissajous orbits allows one the freedom to choose
$\phi_{1}$ and $\phi_{2}$ independently in such a way that \eref{eq:42}
is valid.  We conveniently choose
\begin{equation}
\phi_{1} = \nu_{1}\chi + \frac{\epsilon}{p}, \qquad
\phi_{2} = -\nu_{2}\chi + \frac{\epsilon}{q},
\label{eq:43}
\end{equation}
where $\nu_{1}$ and $\nu_{2}$ are integers and $\epsilon$ is real.
Hence the relation \eref{eq:42} now becomes 
\begin{equation}
(p\nu_{1}+q\nu_{2})\chi = \phi + 2\pi k.
\label{eq:44}
\end{equation}
We now invoke the Bezout's identity  \cite{bezout} which states that
there exist integers 
$\nu_{1}$ and $\nu_{2}$ such that one can always express the greatest
common divisor of $p$ and $q$ ($M$ say) in the form
$p\nu_{1}+q\nu_{2} = M$. Hence it follows from Bezout's identity that
if we choose $\nu_{1}$ and $\nu_{2}$ such that 
$p\nu_{1}+q\nu_{2} = M$ then the relation \eref{eq:44} becomes 
\begin{equation}
\chi = \frac{\phi}{M} + \frac{2\pi k}{M}, \quad
k=0,1,\ldots, M-1.
\label{eq:45}
\end{equation}
If one assumes that $p$ and $q$ are relatively
prime (\ie, $M=1$), then $k=0$ is the only possibility in \eref{eq:42}
and we thus get the unique solution
\begin{equation}
p\phi_{1}-q\phi_{2} = \phi.
\label{eq:46}
\end{equation}
On the other hand if $p$ and $q$ have a common factor $M$, then
in view of \eref{eq:45} one gets the solution,
\begin{equation}
p\phi_{1}-q\phi_{2} = \phi + 2\pi k, \quad k=0,1,\ldots, M-1. 
\label{eq:47}
\end{equation}

Note that the expressions given in \eref{eq:41} for the amplitudes
and in \eref{eq:46} and \eref{eq:47} for the phases, in the case
when $p$ and $q$ are
coprime and not coprime respectively, when 
substituted in the equations for the Lissajous orbit \eref{eq:9},
in the $N \gg 1$ limit, (and for $\phi_{2}=0$), agree \cite{makowski-footnote}
with the solutions \eref{eq:18}, \eref{eq:19}, \eref{eq:20}
guessed by Chen and coworkers \cite{chen-expt,chen-huang1}, based on
their numerical study of the coordinate space probability
densities associated with the coherent state \eref{eq:27}.
Besides as noted by
these authors these solutions also agree with the experimental results
\cite{chen-expt}. Hence this agreement provides an \textit{a posteriori}
justification for our prescription \eref{eq:35} for effecting the
classical limit in this problem.

\section{Conclusions}

In this paper we have exploited the canonical transformation (given by
Louck \etal \cite{louck}) from the commensurate anisotropic oscillator to the
isotropic oscillator in order to construct appropriate $SU(2)$
coherent states for the commensurate anisotropic oscillator over
a degenerate eigenspace. We have demonstrated the classical limit
via the expectation values of the underlying generators.
We have derived explicit expressions for the parameters in the
Lissajous orbit equations in terms of the parameters
of the $SU(2)$ coherent state.
In particular our work completely accounts for
the fact that the SU(2) coherent state in general corresponds to an
ensemble of Lissajous orbits.

It will be interesting to extend the procedure employed in the present
paper to the case of commensurate two-dimensional anisotropic
oscillator in the presence of a weak nonlinear coupling
\cite{chen-nonlin} and the three-dimensional commensurate anisotropic
oscillator \cite{chen-3d} both of which have been experimentally
investigated using analog optical systems recently. We hope to address
these questions in our future work.



\begin{thebibliography}{100}
\bibitem{louck-footnote} Note that although Louck \etal \cite{louck} have
considered $q$ and $p$ to be coprime, their analysis goes through even
if one considers them to be having a common factor.
\bibitem{makowski-footnote} As a matter of fact, our expressions
\eref{eq:41} for the amplitudes completely agree with the `corrected'
expressions conjectured by Makowski \cite{makowski} based on his exact
analysis of the $p=q=1$, $\tau=\pm i$ case.
\end{thebibliography}
\end{document}